\documentclass[a4paper,twocolumn,10pt]{quantumarticle}
\pdfoutput=1

\usepackage{xcolor}
\usepackage[pdftex]{graphicx}
\usepackage{microtype}
\usepackage{hyperref}
\usepackage[braket, qm]{qcircuit}
\usepackage{graphicx}
\usepackage[cmex10]{amsmath}
\usepackage{array}
\usepackage{url}
\usepackage{braket}
\usepackage{amssymb}
\usepackage{caption}
\usepackage{tikz}
\usepackage{adjustbox}
\usepackage{subcaption}
\usepackage{tcolorbox}
\usepackage[cachedir=.]{minted}
\usepackage{listings}
\usepackage{array}
\setminted[python]{
fontsize = \footnotesize,
}

\tcbset{colback=black!5!white,arc=2pt,boxrule=0.5pt}
\BeforeBeginEnvironment{minted}{\begin{tcolorbox}}
\AfterEndEnvironment{minted}{\end{tcolorbox}}

\begin{document}

\title{Methods and Tools for Secure Quantum Clouds with a specific Case Study on Homomorphic Encryption}

\author{Aurelia Kusumastuti}
\affiliation{Technical University of Berlin, Germany}
\email{kusumastuti@campus.tu-berlin.de}

\author{Nikolay Tcholtchev}
\affiliation{RheinMain University of Applied Sciences, Wiesbaden, Germany}
\affiliation{Fraunhofer Institute for Open Communication Systems (FOKUS), Berlin, Germany}
\email{nikolay.tcholtchev@hs-rm.de}
\email{nikolay.tcholtchev@fokus.fraunhofer.de}

\author{Philipp Lämmel}
\affiliation{Fraunhofer Institute for Open Communication Systems (FOKUS), Berlin, Germany}
\email{philipp.laemmel@fokus.fraunhofer.de}

\author{Sebastian Bock}
\email{sebastian.bock@fokus.fraunhofer.de}
\affiliation{Fraunhofer Institute for Open Communication Systems (FOKUS), Berlin, Germany}

\author{Manfred Hauswirth}
\affiliation{Fraunhofer Institute for Open Communication Systems (FOKUS), Berlin, Germany}
\affiliation{Technical University of Berlin, Germany}
\begin{abstract}
  The rise of quantum computing/technology potentially introduces significant security challenges to cloud computing, necessitating quantum-resistant encryption strategies as well as protection schemes and methods for cloud infrastructures offering quantum computing time and services (i.e. quantum clouds). This research explores various options for securing quantum clouds and ensuring privacy, especially focussing on the integration of homomorphic encryption (HE) into Eclipse Qrisp, a high-level quantum computing framework, to enhance the security of quantum cloud platforms. The study addresses the technical feasibility of integrating HE with Qrisp, evaluates performance trade-offs, and assesses the potential impact on future quantum cloud architectures. The successful implementation and Qrisp integration of three post-quantum cryptographic (PQC) algorithms demonstrates the feasibility of integrating HE with quantum computing frameworks. The findings indicate that while the Quantum One-Time Pad (QOTP) offers simplicity and low overhead, other algorithms like Chen and Gentry-Sahai-Waters (GSW) present performance trade-offs in terms of runtime and memory consumption. The study results in an overall set of recommendations for securing quantum clouds, e.g. implementing HE at data storage and processing levels, developing Quantum Key Distribution (QKD), and enforcing stringent access control and authentication mechanisms as well as participating in PQC standardization efforts.
\end{abstract}

\maketitle
\vspace{-0.3em}
\section{Introduction}
\label{sec:introduction}
Cloud computing is a concept that allows for widespread, convenient, and on-demand access to a shared pool of configurable computing resources such as networks, servers, storage, applications, and services, which can be quickly provisioned and released with minimal management effort or interaction with the service provider\cite{mell_nist_2011}. The rapid progression of quantum computing is poised to introduce quantum cloud services to offer a feasible way for ordinary users to use quantum computers, since technological limitations render quantum personal computers at this time unrealistic \cite{fellous_limitations_2021}. 

However, the decentralized nature of cloud services and the storage of user data in the cloud introduce a range of security concerns. These concerns include data breaches, service interruptions, and specific threats like denial-of-service attacks and malware \cite{alouffi_cloud_2021}.  Furthermore, the anticipated capability of quantum computers to compromise traditional cryptographic defenses \textemdash e.g. with Shor's algorithm \cite{bhatia_shor_2020} \textemdash calls for the exploration of quantum-resistant encryption strategies. Among these, certain homomorphic encryption (HE) techniques are recognized for their quantum-resistance, offering a strategy to protect quantum clouds from "store now, decrypt later" attacks and secure data for the future. "Store now, decrypt later" attacks involve adversaries storing encrypted data with the intention of decrypting it in the future, when they have access to more advanced decryption methods or computational power.

\subsection{Technical Research Questions}

The technical research questions of this article are centered on exploring the integration of Eclipse Qrisp, a high-level quantum computing framework developed in Python \cite{seidel_qrisp_2022}, with HE in the context of quantum cloud platforms. Specifically, this article addresses:

\begin{itemize}
    \item \textbf{Integration Challenges}: Investigating the core challenges associated with integrating HE into the Eclipse Qrisp framework. This includes identifying and overcoming the technical hurdles that arise due to the inherent data type incompatibilities between Eclipse Qrisp and the existing libraries for HE.
    \item \textbf{Technical Feasibility}: Assessing the technical feasibility of adapting HE to work seamlessly within the Eclipse Qrisp environment. This involves exploring various strategies to modify HE functionalities to be compatible with Eclipse Qrisp without compromising their security features.
    \item \textbf{Impact on Future Quantum Cloud Platforms}: Evaluating how the successful integration of HE into Eclipse Qrisp could transform the operational capabilities of quantum cloud platforms, potentially leading to more secure quantum computing services.
\end{itemize}

Beyond the above stated technical research questions, the current study conducts a comprehensive state-of-the-art
research regarding protection mechanisms for cloud services and ellucidates on those from the perspective of quantum cloud provider. This leads to a set of general recommendations, which - together with the deep HE technical research activities - set a framework on, how future quantum clouds should be protected with respect
to privacy and cybersecurity issues.

\subsection{Contributions}

The contributions of this article are as follows:
\begin{itemize}
    \item \textbf{Algorithm Assessment}: We outline and assess the feasibility of algorithms that facilitate the implementation of quantum-safe HE within the Eclipse Qrisp framework. These algorithms aim to bridge the gap caused by data type incompatibilities and enable integration.
    \item \textbf{Architectural Design}: We propose an architecture for a secure quantum cloud that incorporates Eclipse Qrisp. This architecture is designed to leverage the strengths of quantum computing while ensuring data security using advanced encryption methods.
    \item \textbf{Implementation of PQC Algorithms}: We detail the implementation of HE addition for three viable PQC algorithms within the Eclipse Qrisp framework:
    \begin{itemize}
        \item A HE scheme \cite{chen_homomorphic_2023} based on McEliece cryptography \cite{mceliecepublic}.
        \item A HE scheme \cite{gsw_2013} based on the Learning With Errors (LWE) problem \cite{regev_lwe_2009}.
        \item A HE scheme \cite{fisher_quantum_2014} using one-time pads.
    \end{itemize}
		\item \textbf{Cybersecurity Recommendations for Quantum Clouds}: We analyse current best practices in the domain of cloud security and formulate recommendations for secure and privacy protecting quantum clouds.
\end{itemize}
\section{Background and related work}

This section provides an overview of foundational concepts and recent advancements in quantum computing, cloud security, post-quantum cryptography, and homomorphic encryption, thus forming the essential basis for this research.

\subsection{Quantum computing}

Quantum computing leverages principles of quantum mechanics, such as superposition and entanglement, to perform data operations, offering a fundamentally different approach from classical computing. Unlike classical bits, which can be either 0 or 1, quantum bits (qubits) can exist in multiple states simultaneously. For instance, a 2-qubit system can represent a superposition of all four possible states ($\ket{00}$, $\ket{01}$, $\ket{10}$, 
$\ket{11}$) at once, exponentially increasing computational capacity as more qubits are added \cite{meng_review_2020}.

Qubits can also be entangled, meaning their states are interdependent regardless of physical separation. Manipulating one entangled qubit instantaneously affects its partner, a property that is crucial to achieve quantum advantage, e.g, in quantum simulations and other advanced applications \cite{meng_review_2020}. However, building a quantum computer faces significant challenges, such as sensitivity to noise, which introduces errors in quantum algorithms. Current fault-tolerant quantum computing models assume constant error rates, but physical error rates tend to increase with scale, thus further complicating building large scale fault-tolerant quantum computers \cite{deleon_noise_2021, fellous_limitations_2021}.

Present-day quantum computers, known as Noisy Intermediate-Scale Quantum (NISQ) devices, lack full error correction, resulting in coherence loss and operational errors. Examples of such devices include IBM's 433-qubit Osprey processor \cite{choi_osprey_2022} and Quantinuums System model H2 \cite{Quantinuum_H2}. On the software side, frameworks like Qiskit \cite{wille_qiskit_2019}, Cirq \cite{garcía_systematic_2023}, and Qrisp \cite{seidel_qrisp_2022} facilitate quantum algorithm development by abstracting the complexities of quantum mechanics.

Eclipse Qrisp, an embedded domain-specific language in Python, simplifies quantum algorithm development by using high-level abstractions. It replaces gates and qubits with functions and variables, allowing for the creation of intricate circuits and easier implementation compared to low-level approaches \cite{seidel_qrisp_2024}. Key features include the `QuantumVariable` class, which manages qubit operations and types like `QuantumFloat`, `QuantumBool`, `QuantumChar`, and `QuantumModulus`. The `QuantumSession` automates lifecycle management and session merging, further simplifying the efficient implementaiton of complex quantum algorithms. Qrisp includes implementations of major quantum algorithms, such as Shor's and Grover's algorithm, although it faces challenges like reliance on Python, which limits execution speed, and complexity in managing high-level abstractions \cite{seidel_backtracking_2024}. Ongoing efforts aim to address these limitations, including integration with Jax for faster compilation \cite{seidel_qrisp_2024}.

\subsection{Cloud security}

Cloud computing provides on-demand access to configurable resources such as networks, servers, storage, and applications, delivering scalability, cost efficiency, and flexibility that allow businesses to scale operations and pay only for what they use \cite{alouffi_cloud_2021}. However, security remains a primary concern, posing threats to confidentiality, integrity, and availability \cite{goodman_cia_2021}.

Confidentiality in cloud computing is typically ensured through encryption, protecting data from unauthorized access. However, long-term confidentiality is challenged by advancements in computational power, especially quantum computing. Quantum computers can potentially break classical cryptographic methods, necessitating the development of quantum-resistant encryption algorithms to ensure long-term data security \cite{mosca2018cybersecurity}. Authentication and authorization frameworks such as OAuth, OpenID, and Keycloak play crucial roles in maintaining security by managing access to resources and minimizing credential exposure \cite{hardt2012oauth, recordon2006openid, redhat2021keycloak}.

Integrity threats, including data tampering and unauthorized changes, can be mitigated through data integrity checks using cryptographic hash functions and maintaining audit trails that log system activities and transactions \cite{gauravaram_cryptographic_2007}. Availability is ensured by implementing failover capabilities for seamless transitions during system failures and conducting regular backups to restore data in case of outages \cite{alouffi_cloud_2021}.

Advances in cloud security are driven by the need to counter sophisticated cyber threats and prepare for the quantum era. Quantum computing holds the potential to revolutionize cloud resource management but also necessitates quantum-resistant encryption methods \cite{zhu_qkd_2023}. Homomorphic encryption (HE) maintains confidentiality during data processing, and quantum-resistant algorithms are being developed to replace current cryptographic standards \cite{dam_pqc_2023}. Identity and Access Management (IAM) systems now incorporate multi-factor authentication and role-based access control, enhancing security by requiring multiple verification factors and restricting access based on roles \cite{ghaffari_identity_2022}.

Anomaly detection has improved with AI integration, enabling real-time identification and response to security incidents, thereby reducing attack opportunities \cite{wu_aisec_2020}. AI can enhance device authentication, detect DoS and DDoS attacks, identify security breaches, and improve malware detection by analyzing network traffic and identifying malicious behaviors \cite{wu_aisec_2020}. Despite potential negative effects like data privacy concerns and algorithmic bias, strategic approaches such as federated learning and continuous monitoring can mitigate these issues and maintain security effectiveness \cite{liu_ml_2018, wu_aisec_2020}.

\subsection{Post-quantum cryptography}

Classical cryptography relies on the computational difficulty of certain mathematical problems, such as the integer factorization problem for RSA \cite{rivest_rsa_1978}, the elliptic curve discrete logarithm problem for Elliptic Curve Cryptography \cite{miller_ecc_1986}, and the discrete logarithm problem for Diffie-Hellman \cite{diffie_new_2022}. However, quantum computing poses a significant threat to these cryptosystems. Peter Shor's quantum algorithm, introduced in 1994, can solve these problems in polynomial time, rendering classical cryptographic techniques vulnerable \cite{shor_rsa_1994}. This vulnerability has led to the development of post-quantum cryptography (PQC).

PQC focuses on cryptographic algorithms that remain secure even against quantum computers. These algorithms are built on mathematical problems that quantum algorithms do not solve significantly faster than classical ones \cite{dam_pqc_2023}. Key types of PQC algorithms include:

\begin{itemize}
    \item \textbf{Lattice-based cryptography}: Relies on the hardness of lattice problems, such as the Learning With Errors (LWE) problem \cite{richter_pqc_2022}. An example is the Gentry-Sahai-Waters (GSW) encryption \cite{gsw_2013}.
    \item \textbf{Code-based cryptography}: Based on the difficulty of decoding linear codes, exemplified by the McEliece cryptosystem \cite{mceliecepublic}. Chen's work on homomorphic encryption builds on this system \cite{chen_homomorphic_2023}.
    \item \textbf{Multivariate polynomial cryptography}: Involves solving systems of multivariate quadratic equations, which are NP-hard \cite{richter_pqc_2022}.
    \item \textbf{Hash-based cryptography}: Utilizes cryptographic hash functions that are designed to be one-way and collision-resistant \cite{richter_pqc_2022}.
\end{itemize}
    
The National Institute of Standards and Technology (NIST) has been leading efforts to standardize PQC algorithms. NIST's open competition aims to identify cryptographic schemes that can withstand quantum capabilities and be easily integrated into existing systems \cite{Yesina_nist_2022}. The third round of this competition has selected algorithms like CRYSTALS-Kyber, a lattice-based key encapsulation mechanism, and the classic McEliece cryptosystem \cite{bos_kyber_2018, mceliecepublic}.

Challenges in PQC include ensuring a smooth transition from classical to quantum-resistant systems, optimizing algorithms for diverse hardware and software platforms, and addressing specific use cases such as IoT devices with resource constraints \cite{pandey_pqc_2023}. Additionally, comprehensive security proofs and resistance to both quantum and classical attacks, including side-channel attacks, are necessary to ensure robust security \cite{canto_sca_2023}.

\subsection{Homomorphic encryption}

Homomorphic encryption (HE) allows computations on encrypted data without decryption. For any plaintexts $a$ and $b$, an encryption function $enc()$, and a binary operation $+$, the property $enc(a+b)=enc(a)+enc(b)$ holds \cite{logsdon_he_2023}. This is particularly useful in cloud computing and collaborative environments where sensitive data needs processing.

HE comes in several forms: Partially Homomorphic Encryption (PHE), Somewhat Homomorphic Encryption (SHE), and Fully Homomorphic Encryption (FHE). PHE supports either addition or multiplication an unlimited number of times, making it efficient but limited in scope. RSA is an example of PHE for multiplication \cite{chaudhary_phe_2019}. SHE allows both operations but only for a limited number of times due to noise accumulation, which can make ciphertexts indecipherable \cite{logsdon_he_2023, chaudhary_phe_2019}. SHE often serves as a foundation for FHE through a process called bootstrapping, which refreshes the ciphertext to manage noise \cite{gentry_bootstrapping_2009}. FHE supports unlimited addition and multiplication, enabling arbitrary computations on ciphertexts, but is computationally intensive and resource-heavy due to complex noise management \cite{logsdon_he_2023}.

The evolution of FHE has seen significant milestones. Craig Gentry's 2009 scheme introduced lattice-based cryptography and bootstrapping, laying the groundwork for FHE \cite{gentry_bootstrapping_2009}. Although groundbreaking, this first generation was highly inefficient. The second generation, around 2011-2012, introduced leveled FHE, which allowed computations up to a certain depth without frequent bootstrapping, improving practicality \cite{brakerski_efficient_2011, gentry_implementing_2011}. The third generation, between 2013-2014, saw practical schemes like BGV and GSW that used Ring-LWE problems for enhanced security and efficiency \cite{brakerski_leveled_2014, gsw_2013}. Batching techniques and libraries like HElib made these more accessible \cite{halevi_algorithms_2014}. The fourth generation, from 2015 onwards, brought advanced schemes like BFV and CKKS, which support approximate arithmetic, making FHE viable for applications like machine learning \cite{fan_somewhat_2012}. Current efforts focus on standardization and integration with cloud services, aiming for broader adoption and practicality.

Despite advancements, challenges remain in adapting HE for post-quantum cryptography (PQC). Issues include algorithmic efficiency and noise management during homomorphic operations. Research into noise-free schemes and new algorithms with quantum-resistant assumptions is ongoing \cite{mukherjee_lattice_2023, leonardi_noise_2023}.
\section{General Requirements for Ensuring the Security of Quantum Clouds}

In addition to confidentiality, integrity, and availability, the National Institute of Standards and Technology (NIST) identifies authentication, authorization, accountability, and privacy as essential cloud security requirements \cite{liu_nist_2011}. This section discusses specific measures to fulfill these requirements and expands them to include quantum clouds.

Due to the interconnected nature of cloud security, some requirements overlap. For instance, robust authentication mechanisms enhance confidentiality and integrity by ensuring that only authorized users access and modify data. Similarly, accountability measures support privacy by providing traceability and non-repudiation, crucial for compliance with data protection regulations. This overlap ensures a comprehensive and cohesive security framework for quantum clouds.

\subsection{Confidentiality}

Confidentiality protects a cloud user's sensitive data from unauthorized access \cite{liu_nist_2011}. This is crucial in cloud computing due to the risks associated with storing and processing data on remote servers managed by third-party providers.

\subsubsection{Encryption}

Quantum-specific attacks, such as those enabled by Shor's algorithm, threaten classical cryptographic protocols like RSA and Elliptic Curve Cryptography \cite{shor_rsa_1994}. To ensure data confidentiality, integrating quantum-resistant End-to-End Encryption (E2EE) with Homomorphic Encryption (HE) offers robust security. E2EE safeguards data during transmission by encrypting it from the source to the destination, preventing unauthorized access. Concurrently, HE allows secure processing of encrypted data without decryption, maintaining confidentiality throughout the processing phase.

\subsubsection{Secure Key Management}

Beyond implementing PQC, additional measures are essential to ensure confidentiality against quantum threats like quantum eavesdropping. Quantum Key Distribution (QKD) uses quantum mechanics to securely distribute encryption keys. It leverages quantum entanglement, the effect that measuring a quantum system disturbs it, and the No-Cloning Theorem, which asserts the impossibility of creating an exact copy of an unknown quantum state. These principles make eavesdropping detectable.

\begin{figure}[ht]
\centering
\includegraphics[width=0.5\textwidth]{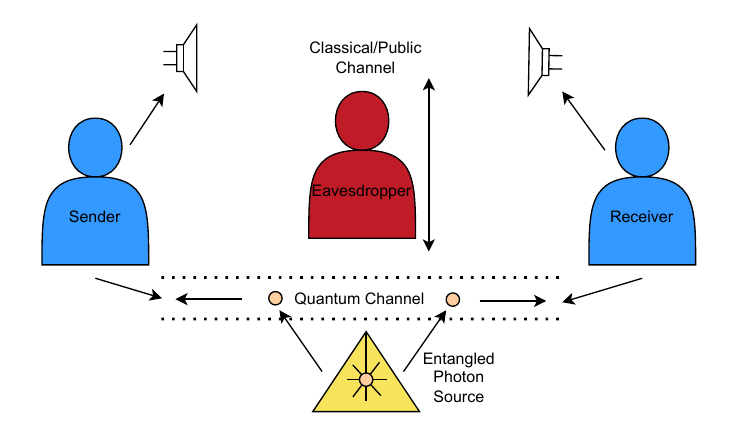}
\caption{Typical QKD communication setup}
\label{fig:qkd}
\end{figure}

A typical quantum communication setup with a sender, receiver, and eavesdropper is depicted in Figure 
\ref{fig:qkd}. An entangled photon source is used to secure the quantum channel. In QKD, the sender could for example send photons encoded with key bits through a quantum channel. The receiver measures the photons using random bases. Both parties then compare a subset of their measurements to detect eavesdropping. If the error rate is low, they generate a secure key.

QKD protocols, such as BB84 and E91, use qubits to encode key information:
\begin{itemize}
	\item BB84 Protocol: Uses the polarization states of photons to represent qubits. The sender transmits photons to the receiver, who measures them with randomly chosen bases. After transmission, they compare their bases publicly and discard mismatched measurements to form a shared secret key \cite{bennett_bb84_2014}.
	\item E91 Protocol: Based on quantum entanglement, pairs of entangled particles are distributed to the sender and receiver. Measurements on these particles are correlated, and any eavesdropping attempt disturbs the entanglement, revealing the intruder \cite{ekert_e91_1991}.
\end{itemize}

In QKD, any attempt to intercept qubits disturbs their quantum state, alerting the sender and receiver to discard the compromised key, thus ensuring secure communication.

\subsection{Data Integrity}

Data integrity guarantees the accuracy, consistency, and reliability of cloud-stored data \cite{liu_nist_2011}.

Cryptographic Hash Functions: These functions produce a fixed-size hash value from input data, which uniquely represents the data. Any change in the input data results in a significantly different hash, making unauthorized modifications easy to detect. Hash functions can verify data integrity in quantum cloud environments by hashing data before and after processing and comparing the hashes \cite{gauravaram_cryptographic_2007}. Quantum-resistant hash functions, like those based on lattice cryptography, provide security against quantum attacks.

Digital Signatures: These authenticate data and verify its integrity by creating a unique signature with the sender's private key, verifiable by anyone with the sender's public key. In quantum clouds, digital signatures ensure that data has not been altered since signing. Quantum-resistant digital signatures, such as twin-field quantum digital signatures based on QKD, are being developed to withstand quantum attacks \cite{Zhang_twinfield_2021}.

\subsubsection{System Integrity}

System integrity ensures that cloud infrastructure, both hardware and software, remains unaltered and operates correctly \cite{liu_nist_2011}.

Secure Firmware and Software Updates: Secure updates should use cryptographic verification to validate updates before application, preventing malicious code introduction. In quantum clouds, both quantum and classical components need robust update verification processes, such as signing updates with a trusted key and verifying the signature before installation \cite{choi_firmware_2016}.

Trustworthy Quantum Hardware: Quantum Physically Unclonable Functions (QuPUFs) provide a unique, tamper-evident fingerprint for each quantum hardware component, ensuring authenticity and integrity. QuPUFs leverage the no-cloning theorem, making them resistant to cloning and various attacks \cite{Arapinis_qupuf_2021}. Cryptographic verification protocols should regularly verify quantum hardware integrity, ensuring only trusted hardware is used for computations. This can involve QuPUF-based signatures verified against known values during the boot process to detect tampering or hardware failure.

\subsection{Availability}

Availability ensures that quantum cloud infrastructure and services remain accessible during hardware failures, software bugs, cyber-attacks, or natural disasters \cite{liu_nist_2011}.

\subsubsection{Redundancy and Failover Systems}

Deploy redundant quantum processing units (QPUs) and classical hardware to ensure service continuity. Redundancy involves having multiple QPUs and servers that can take over workloads if one unit fails, preventing service interruptions. Automated failover systems detect failures and switch to backup resources without manual intervention, ensuring continuous service availability. Regular testing of these systems is essential to ensure functionality during actual failures \cite{britt_qhpc_2017}.

\begin{figure*}
	\centering
	\includegraphics[width=\textwidth]{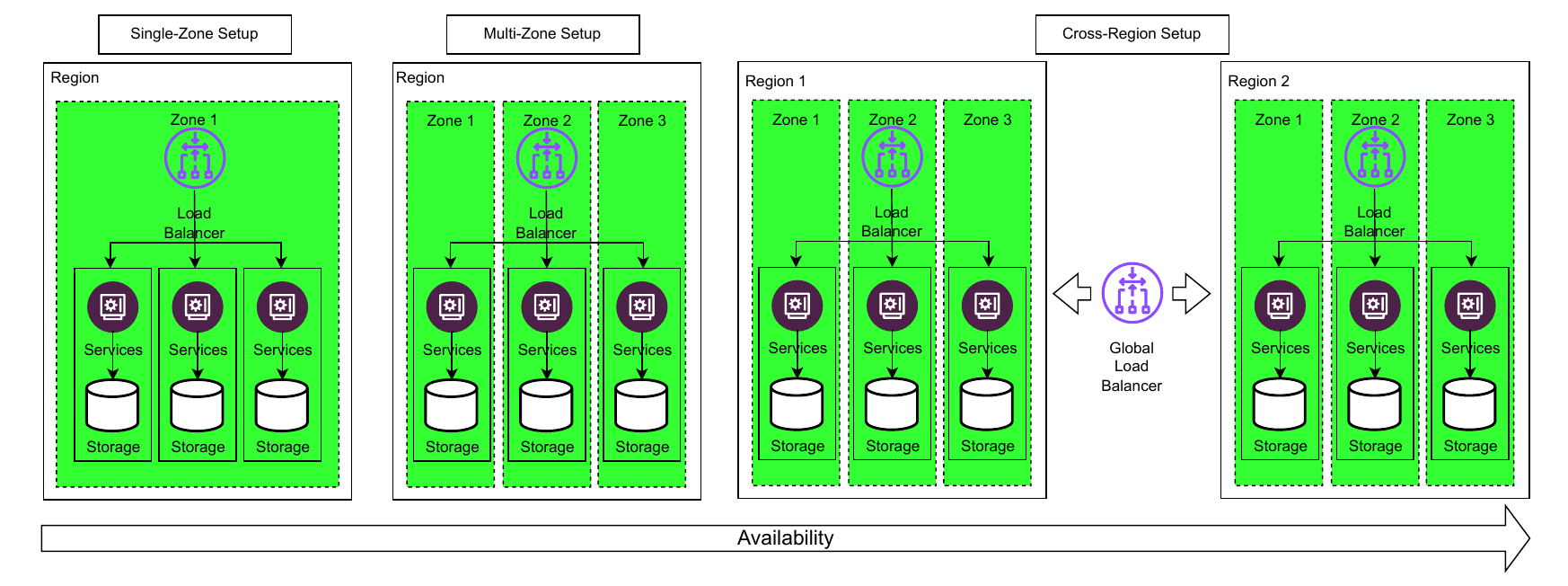}
	\caption{High availability cloud architecture patterns}
	\label{fig:availability}
\end{figure*}

Figure \ref{fig:availability} sketches relevant high-availability cloud architecture patterns.
These include redundant services within one zone, multi-redundant services across multiple zones in one region, and redundant services across different regions. Each setup shows the distribution of services, load balancers, and storage across different zones and regions to improve availability.

\subsubsection{Backup Strategies}

The following measures should be considered to ensure reliable backups in a quantum cloud environment: 
\begin{itemize}
	\item Implement comprehensive backup strategies to ensure data availability during catastrophic events.
	\item Perform regular data backups, storing copies both on-site and off-site to protect against data loss from hardware failures, cyber-attacks, or natural disasters. 
	\item Backup strategies should include quantum data and the configurations and software necessary to restore services. Disaster recovery plans should outline steps to restore services and data, prioritizing critical systems \cite{nguyen_iquantum_2024}.
\end{itemize}

\subsubsection{Scalability}

The following measures can be considered to address scalability concerns in a quantum cloud environment:
\begin{itemize}
	\item Design scalable security measures to handle increased workloads efficiently. 
	\item Ensure that quantum and classical resources can scale to meet demand. 
	\item Implement dynamic resource allocation algorithms to scale the number of active QPUs based on current demand. 
	\item Leverage the  processing capabilities of quantum algorithms to enhance security protocols, such as faster encryption and decryption processes \cite{dumitrescu_nuclei_2018}.
\end{itemize}

\subsubsection{Isolation}

In addition to the above aspects, components and ressources in a quantum cloud should be properly isolated 
and protected within the relevant scope:
\begin{itemize}
	\item Ensure isolation to securely host multiple tenants without interference, preventing attacks such as fault injection and denial-of-service in shared environments. 
	\item Use techniques like virtual private networks (VPNs), dedicated hardware resources, and software-defined networking (SDN) to achieve isolation. 
	\item  Using dedicated QPUs for different tenants can prevent cross-tenant interference, and container-based performance isolation mechanisms can reduce performance competition among tenants \cite{wang_container_2020}.
\end{itemize}

\subsection{Authentication}

Robust authentication mechanisms are crucial for verifying the identity of users and devices accessing quantum cloud services, preventing unauthorized access and ensuring only legitimate entities interact with the cloud infrastructure \cite{liu_nist_2011}.

\subsubsection{Multi-Factor Authentication (MFA)}

Implementing MFA enhances security by requiring multiple verification factors from different categories: something the user knows (e.g., a password), something the user has (e.g., a smartphone or hardware token), and something the user is (e.g., biometric data) \cite{ometov_mfa_2018}. Advanced MFA methods incorporate user behavior profiles and fuzzy hashing to detect impostors effectively while preserving privacy.

\begin{figure}[ht]
	\centering
	\includegraphics[width=0.5\textwidth]{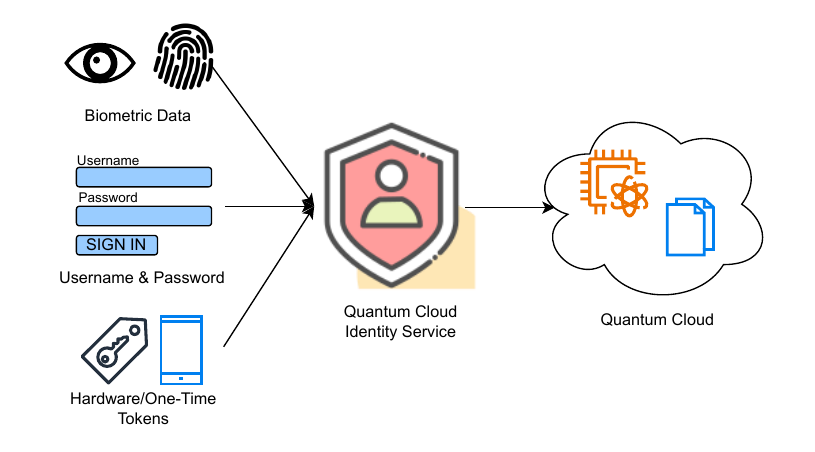}
	\caption{Examples of MFA elements.} 
	\label{fig:auth}
\end{figure}

Figure \ref{fig:auth} shows an example for MFA elements embedded in a cloud architecture. 
The goal is to provide a quantum cloud-based multi-factor authentication system, which should be in place
for securing quantum clouds. In this scope, users provide biometric data, username, password, or hardware/one-time tokens to sign in. These credentials are verified by the quantum cloud identity service, which uses quantum technology to enhance security.

\subsubsection{Secure Communication Channels}

Using encrypted communication protocols like Transport Layer Security (TLS) protects authentication credentials during transmission. These protocols encrypt data in transit, preventing interception by attackers \cite{oppliger_ssl_2023}. To mitigate known vulnerabilities in SSL/TLS, it is crucial to disable outdated versions, use strong cipher suites, and implement proper certificate validation. Integrating post-quantum key exchange methods with traditional digital signatures in TLS enhances security against quantum threats \cite{bos_pqtls_2015}.

\subsubsection{Digital Certificate Management}

Effective management of digital certificates involves issuing, renewing, and revoking certificates to maintain trust and security. Digital certificates authenticate the identity of entities and establish secure communications. For instance, quantum cloud providers can use digital certificates to authenticate connections between quantum computing nodes and classical control systems. Quantum-resistant implementations, such as DOPIV \cite{zhang_dopiv_2022} and post-quantum digital signatures like Dilithium \cite{ducas_crystals_2018} and Falcon \cite{fouque_falcon_2018}, integrated into public key certificates, enhance security against quantum threats.

\subsection{Authorization}

Effective authorization ensures that only authorized users and applications access sensitive data and resources, mitigating the risk of unauthorized access and potential data breaches. Stringent authorization protocols enhance security and ensure compliance with regulatory requirements in the cloud \cite{liu_nist_2011}.

\subsubsection{Role-Based Access Control (RBAC)}

\begin{figure}[htb]
	\centering
	\includegraphics[width=0.5\textwidth]{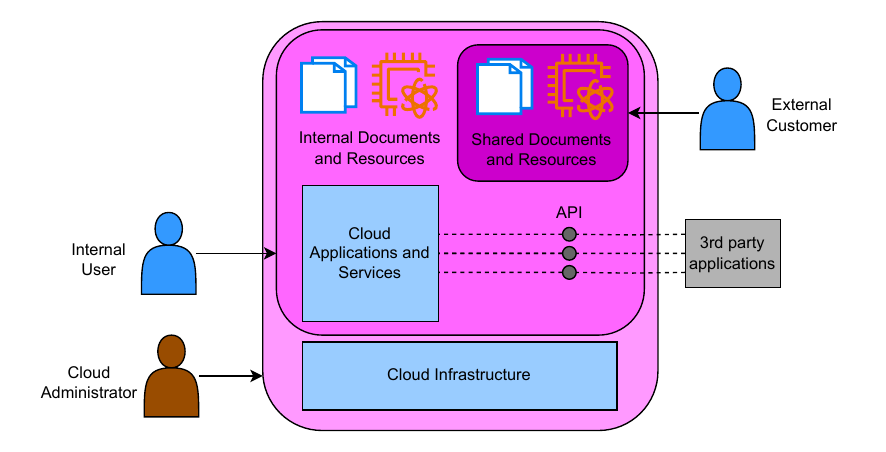}
	\caption{Authorization boundaries for administrator, internal user, and external customer permissions in a quantum cloud system.}
	\label{fig:rbac}
\end{figure}

RBAC restricts system access based on user roles within an organization. 
This especially relates to cloud computing architectures with internal and external users accessing cloud applications and services as illustrated in Figure \ref{fig:rbac}. The cloud administrator manages the cloud infrastructure, providing access to internal and shared documents and resources. Third-party applications can also interact with the cloud environment through an API.

Within RBAC, permissions are assigned to roles rather than users, and users are then assigned roles, simplifying permission management. For example, a "Quantum Researcher" might access quantum algorithms, while a "System Administrator" manages the quantum cloud infrastructure.

Implementing fine-grained permissions within the RBAC framework ensures users have only the necessary permissions for their roles \cite{zhu_finerbac_2011}. Access to QPUs, and quantum key distribution channels must be meticulously controlled to prevent unauthorized manipulation or observation.

\subsubsection{Policy Generation and Enforcement}

Users should have the minimum permissions necessary to perform their tasks, minimizing the attack surface and potential damage from compromised accounts. Regular audits and reviews of user permissions ensure they remain appropriate and aligned with users' evolving roles.

Automated methods for creating least privilege policies from audit logs reduce over-privileging and administrative burden. Frameworks that automatically generate policies from audit log data enhance security by ensuring permissions are limited to the minimum necessary \cite{sanders_autorbac_2017}. In quantum cloud environments, this involves carefully managing access to quantum computation tasks and communication channels.

\subsection{Privacy}

As organizations increasingly rely on public clouds, they face significant privacy challenges, including safeguarding data confidentiality and integrity, complying with diverse regulations, and mitigating unauthorized access and data breaches \cite{liu_nist_2011}.

\subsubsection{Privacy-Preserving Methods}

In the scope of privacy-preserving methods, it is paramount to implement data anonymization and masking techniques to protect sensitive information within quantum cloud infrastructures. These methods alter data to prevent the identification of individuals while preserving its utility. In quantum clouds, anonymization can be applied to datasets used in quantum machine learning, ensuring personal information remains secure \cite{andrew_privacy_2019}. Masking sensitive information during processing further mitigates data breach risks, enhancing privacy protection. Quantum homomorphic encryption (QHE) also enables computations on encrypted data without decryption, preserving privacy throughout the computation process \cite{fisher_quantum_2014}.

\subsubsection{Regulatory Compliance}

Complying with data protection regulations, such as the General Data Protection Regulation (GDPR), is crucial for safeguarding privacy in cloud environments. Quantum cloud providers must implement data handling and protection measures tailored to quantum computing's unique context. This includes obtaining explicit consent for data processing, minimizing data collection, and providing users with rights to access, correct, and delete their data \cite{voigt_gdpr_2017}.
\section{Concepts}

This section addresses secure data processing in quantum computing environments, ensuring sensitive information remains protected even when handled by potentially untrusted quantum servers. We describe the cryptosystems explored in this research, covering their mathematical foundations, key generation, encryption, decryption, and additive homomorphic properties. Additionally, we outline possible quantum cloud applications of these cryptosystems and their integration with Qrisp, offering insights into enhancing security in quantum cloud environments and providing a stepping stone for robust data protection in the post-quantum era.

\subsection{Chen (McEliece Variant)}
\label{sec:concept:chenMcEliece} 

As mentioned in the previous chapters, one of the approaches used for PQC is the Classic McEliece. It is based on the difficulty of decoding a general linear code, which is NP-hard, providing a robust framework for asymmetric encryption. In 2023, Chen \cite{chen_homomorphic_2023} proposed a PHE method for HE binary addition based on the McEliece cryptosystem using Hamming codes as the linear code. Hamming code is an error-correcting code that was developed by Richard W. Hamming in 1950. It is designed to detect and correct single-bit errors in data transmission or storage, making it particularly useful in digital communication systems and computer memory architectures \cite{hamming_code_1950}. The parameters of a Hamming code, namely $k, n$, and $N$ , are defined as follows:

\begin{enumerate}
    \item \textbf{$N$ - Total Number of Bits in the Codeword}: This includes both the data bits and the parity bits. It has the form $N = 2^k - 1$, where $k$ is the number of parity bits. The formula arises because each parity bit covers positions that are powers of two, and each of these positions must be able to index the total length of the codeword.
    \item \textbf{$n$ - Number of Message Bits}: This is the number of bits in the original message that needs to be encoded. It is defined as $n = 2^k - k - 1 = N-k$
    \item \textbf{$k$ - Number of Parity Bits}: The number of parity bits needed for the length of the data bits and the total codeword length.
\end{enumerate}

These parameters ensure that each bit position in the codeword (including the parity bits themselves) can be uniquely checked by a combination of parity bits, allowing for the detection and correction of single-bit errors. 

\paragraph{Key generation.} \label{par:chen_key} In the context of Hamming code as discussed in Chen's cryptosystem, several matrices are used to facilitate the encoding and decoding processes:

\begin{enumerate}
    \item \textbf{Scrambler Matrix ($S$)}: A random invertible $n \times n$ binary matrix. 
    \item \textbf{Generator Matrix ($G$)}: A $n \times N$ matrix constructed as follows:
    \begin{enumerate}
        \item From a $n \times n$ identity matrix $I_{n}$, swap the ones and the zeroes to form the $n \times n$ matrix $I_{n}'$.
        \item Randomly pick $k$ columns from $I_{n}'$.
        \item Construct $G$ by randomly ordering the column vectors selected in step 2 and the columns in $I_{n}$.
    \end{enumerate}
    \item \textbf{Parity-check Matrix ($R$)}: A $N \times n$ matrix which satisfies $GR^{T} = I$.
    \item \textbf{Permutation Matrix ($P$)}: A $N \times N$ identity matrix whose columns have been scrambled.
    \item \textbf{Encryption Matrix ($\Psi$)}: A $n \times N$ matrix, defined as $\Psi = SGP$.
\end{enumerate}

\paragraph{Encryption.}  To encode $n$-bit long binary summands $x_{1}$ and $x_{2}$, we just multiply them by $\Psi$ to get $c_{x1} = Enc(x_{1})= x_{1}\Psi$ and $c_{x2} = Enc(x_{2})= x_{2}\Psi$ respectively. 

\paragraph{Decryption.} After performing the addition $\Phi = c_{x1} \oplus c_{x2}$, we can decrypt $\Phi$ by calculating $\Theta = Dec(\Phi) = \Phi P^{-1}RS^{-1} = x_{1} \oplus x_{2}$.

We can see that the encryption scheme is homomorphic by showing the following:

\begin{align}\label{eq:heChen}
    \begin{split}
        \Phi &= \sum_{i=1}^{n} c_{xi}
        \\&= c_{x1} \oplus c_{x2} \oplus \cdots \oplus c_{xn}
        \\&= x_{1}\Psi \oplus x_{2}\Psi \oplus \cdots \oplus x_{n}\Psi
        \\&= (x_{1} \oplus x_{2} \oplus \cdots \oplus x_{n})\Psi
        \\&= \Theta\Psi = Enc(\Theta)
    \end{split}
\end{align}

\noindent In Equation (\ref{eq:heChen}), we can see that the sum of the encrypted summands equal the encrypted sum of the plaintext summands. Notice however, that Chen's method is homomorphic for parity addition (bitwise XOR), not arithmetic addition. To support arithmetic addition, the carry is calculated using bitwise AND and left shifts before encryption, preserving the homomorphic property. This extension allows correct arithmetic results even when summands have overlapping ones in their binary representations.

\subsection{Gentry-Sahai-Waters}
\label{ch:concept:gsw} 
The Gentry-Sahai-Waters (GSW) cryptosystem is a homomorphic encryption which utilizes lattice-based cryptography, specifically leveraging problems like Learning With Errors (LWE) and its variant Ring-LWE \cite{gsw_2013}. 
LWE is centered around distinguishing between "noisy" linear equations and completely random values. Several key notations are introduced to define LWE:

\begin{itemize}
    \item \textbf{$n$ - Dimension of the secret vector and public matrix}: a positive integer.
     \item \textbf{$\Lambda$ - $n$-dimensional lattice}: a discrete, periodic set of points in $n$-dimensional space with a basis $B = (b_{1},\cdots,b_{n})$.
     \item \textbf{$q$ - Noise modulus}: a positive integer.
     \item \textbf{$s$ - Secret vector}: an unknown $n$-dimensional vector in $\Lambda$.
     \item \textbf{$A$ - Public matrix}: a $m\times n$ matrix ($m$ being the number of equations) with uniformly random integer entries in $[0, q-1]$.
     \item \textbf{$e$ - Error vector}: random vector in $\mathbb{Z}^{m}$ which serves as small errors for each equation in $A$. Each element is drawn from a discrete Gaussian distribution centered at 0.
\end{itemize}

Given a public matrix $A$ and a vector $b = As + e$ where $s$ is the secret vector and $e$ is the error vector as defined above, the goal of (Search-)LWE is to find $s$ \cite{regev_lwe_2009}. Ring-LWE is a similar problem, but instead of operating on vectors over integers, Ring-LWE operates on polynomials over a finite field \cite{lyubashevsky_rlwe_2013}.

In this article, we implement a construction of GSW similar to the one outlined by Minelli \cite{minelli_gsw_2018}.

\paragraph{Key generation.} \label{par:gsw_keys} We choose an integer $n$ to be our security parameter and a modulus $q$ which is $n$ bits long. Specifically, we choose $q$ to be a Sophie-Germain prime, which means not only $q$ is a prime, but also $2q+1$. $2q+1$ is then called the safe prime of $q$. Using Sophie-Germain primes reduces the risk of certain algebraic attacks that exploit special factorizations, making the scheme more secure against these types of attacks \cite{nejatollahi_pqclattice_2019}.

Let $l = \lceil log(q) \rceil $ and $m = nl$. We sample a secret vector $\boldsymbol{s} \leftarrow \mathbb{Z} ^{n-1}_{q}$, a matrix $\boldsymbol{A} \leftarrow \mathbb{Z} ^{(n-1) \times m}_{q}$, and an error vector $\boldsymbol{e} \leftarrow \chi^{m}$, where $\chi$ is a normal distribution of integers modulo $q$. We then have $sk = \boldsymbol{\hat{s}} := (s\|1)$ and the public key $ pk = 
\left(
\begin{array}{c}
-\boldsymbol{A} \\
\boldsymbol{s}^{T}\boldsymbol{A} + \boldsymbol{e}^{T} 
\end{array}
\right) $ \cite{minelli_gsw_2018}.

\paragraph{Encryption.} We construct the following generator matrix: \\ \\
$\boldsymbol{G}^{n \times m} = \\$
\begin{equation*}
\left[
\begin{array}{ccccccccc}
1      & 2      & \cdots & 2^{l-1} & \cdots & 0      & 0      & 0      & 0      \\
0      & 0      & \cdots & 0       & \cdots & 0      & \cdots & 0      & 0      \\
\vdots & \vdots & \vdots & \vdots  & \vdots & \vdots & \vdots & \vdots & \vdots \\ 
0      & 0      & \cdots & 0       & \cdots & 1      & 2      & \cdots & 2^{l-1}\\
\end{array}
\right]
\end{equation*}

Given a message $\mu$, we sample a random matrix $\boldsymbol{R} \leftarrow \{0,1\}^{m \times m}$ and calculate the ciphertext:

\begin{equation} \label{eq:gsw_enc}
    \boldsymbol{C} = \left(
\begin{array}{c}
-\boldsymbol{A} \\
\boldsymbol{s}^{T}\boldsymbol{A} + \boldsymbol{e}^{T} 
\end{array}
\right)\boldsymbol{R} + \mu\boldsymbol{G} \in \mathbb{Z}^{m \times n}_{q}
\end{equation}

\paragraph{Decryption.} Given a ciphertext $C$, we compute

\begin{equation} \label{eq:gsw_dec}
    \hat{s}^{T}\boldsymbol{C} = \hat{s}^{T}\left(\left(
\begin{array}{c}
-\boldsymbol{A} \\
\boldsymbol{s}^{T}\boldsymbol{A} + \boldsymbol{e}^{T} 
\end{array}
\right)\boldsymbol{R} + \mu\boldsymbol{G} \right) = \boldsymbol{e}^{T}\boldsymbol{R}+\mu \hat{s}^{T}\boldsymbol{G} 
\end{equation}
To get the plaintext, we need to retrieve $\mu'$, the closest integer to $\hat{s}^{T}\boldsymbol{G}$ \cite{minelli_gsw_2018}. To calculate $\mu'$, we ignore the small error $\boldsymbol{e}^{T}\boldsymbol{R}$ and calculate the distance $s^{T}\boldsymbol{G}$ to $\hat{s}^{T}\boldsymbol{C}$, namely $d = \|s^{T}\boldsymbol{G} - \hat{s}^{T}\boldsymbol{C}\|$. We choose the value in $s^{T}\boldsymbol{G}$ which has the lowest distance to $\hat{s}^{T}\boldsymbol{C}$ to decrypt to.

To show the homomorphic addition, let $\boldsymbol{C}^{+} := \boldsymbol{C}_{1} + \boldsymbol{C}_{2}$ be the addition of ciphertexts $\boldsymbol{C}_{1}$ and $\boldsymbol{C}_{2}$, each encrypting $\mu_{1}$ and $\mu_{2}$ respectively. From Equation (\ref{eq:gsw_dec}) we expand as follows:

\begin{equation}
    \hat{s}^{T}\boldsymbol{C}^{+} = \hat{s}^{T}\left(\boldsymbol{C}_{1} + \boldsymbol{C}_{2}\right) = \left(\boldsymbol{e}_{1}^{T} + \boldsymbol{e}_{2}^{T}\right) + \left(\mu_{1}^{T} + \mu_{2}^{T}\right) \hat{s}^{T}\boldsymbol{G}
\end{equation}

The above equation shows that $\boldsymbol{C}_{1} + \boldsymbol{C}_{2}$ is the encryption of $\mu_{1} + \mu_{2}$ given that the error terms are negligible.

\subsection{Quantum One-Time Pad (QOTP)}
The encryption scheme proposed by Fisher et al. \cite{fisher_quantum_2014} uses the principles of a one-time pad (OTP) to secure qubits against unauthorized access. OTP, when correctly implemented, is considered theoretically unbreakable due to its perfect secrecy. This is achieved by ensuring that the encryption key is truly random and is never reused. The protocol is therefore able to make encrypted states indistinguishable from random states without the decryption keys.

\paragraph{Key generation.} The key generation process involves creating two classical bits for each qubit that will be encrypted. These bits determine whether a Pauli $\mathbf{X}$ (bit flip) and/or $\mathbf{Z}$ (phase flip) operation is applied to the qubit. For example, bits '01' means that the qubit will be encrypted with no bit flip and a phase flip. For the use case of adding two $n$-qubit integers, we need two lists of classical bits of size $n$ for each integer: one for determining the application of $\mathbf{X}$ gates and the other for the application $\mathbf{Z}$ gates. We will denote these lists to be $a$ and $b$ for the first summand and $c$ and $d$ for the second. The randomness of these bits is crucial for the security of the encryption, as it ensures that the encrypted state is completely mixed and indistinguishable from a random state to anyone without the keys.

\paragraph{Encryption.} We apply the Pauli X and/or Z operations to each qubit of the summands based on the generated keys: the classical bits in $a$ and $b$ encrypt the first summand, $c$ and $d$ encrypt the second summand. Let the first summand be $m_{1}$ and the second $m_{2}$:

\begin{equation} \label{eq:otp_enc_1}
    \mathbf{X}^{a[i]}\mathbf{Z}^{b[i]}\ket{m_{1}[i]} := Enc(\ket{m_{1}[i]})
\end{equation}
\begin{equation} \label{eq:otp_enc_2}
    \mathbf{X}^{c[i]}\mathbf{Z}^{d[i]}\ket{m_{2}[i]} := Enc(\ket{m_{2}[i]})
\end{equation}

$a[i],b[i],c[i],$ and $d[i]$ denote the $i-$th member of $a,b,c,$ and $d$, and $\ket{m_{1}[i]}$ and $\ket{m_{2}[i]}$ denote the $i-$th qubit of $m_{1}$ and $m_{2}$ respectively.

\paragraph{Decryption.} After applying encrypted parity addition using the CNOT operator on the summands, we will get the encrypted result $\ket{Enc(m_{1}), Enc(m_{2})} \mapsto \ket{Enc(m_{1}), Enc(m_{1}) \oplus Enc(m_{2})}$. Since we are only interested in the result of the parity addition, we will only need to decrypt the second term. Let $m_{3} := m_{1} \oplus m_{2}$ and $Enc(m_{3}) := Enc(m_{1}) \oplus Enc(m_{2})$ then:

\begin{equation}
    \ket{m_{3}[i]} := \mathbf{X}^{a[i] \oplus c[i]}\mathbf{Z}^{d[i]}\ket{Enc(m_{3}[i])} 
\end{equation}

As shown by Fisher et al. \cite{fisher_quantum_2014}, the above protocol enables parity addition on encrypted qubits. To support arithmetic addition, we perform a similar enhancement as we did on Chen's protocol: We calculate the carry in advance and add it to the parity addition result at the end. To calculate the carry, we apply the Toffoli gate:

\begin{equation}
    \ket{m_{1}, m_{2}, 0} \mapsto \ket{m_{1}, m_{2}, 0 \oplus (m_{1} \wedge m_{2})}
\end{equation}

Notice that $0 \oplus (m_{1} \wedge m_{2}) = m_{1} \wedge m_{2}$. We then apply one bit shift to the left to get the overall carry.

\subsection{Possible Quantum Cloud Architectures} 

Having discussed the cryptosystems in question, we now outline their implementation in a quantum cloud environment. 

\begin{figure}[ht]
	\centering
	\includegraphics[width=0.5\textwidth]{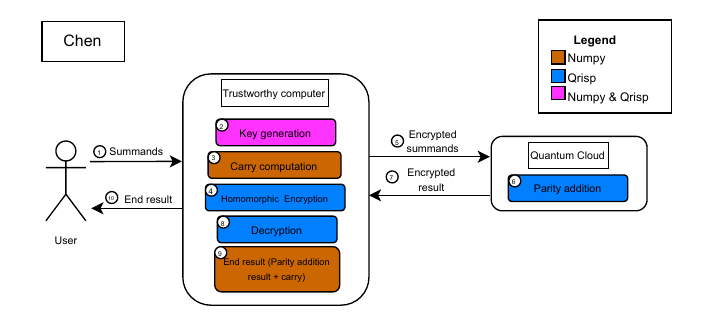}
	\caption{Example architecture of quantum cloud usage for homomorphic addition using the Chen algorithm.}
	\label{fig:chen_arc}
\end{figure}

\paragraph{Chen.} Using the homomorphic encryption introduced by Chen \cite{chen_homomorphic_2023}, the user defines the summands and calculates the carry on a trusted computer. The matrices $S, G, P, R, P^{-1}$, and $S^{-1}$ are generated for encryption and decryption. The summands are encrypted using $\Psi = SGP$ and sent to the quantum cloud for processing. The cloud performs a homomorphic parity addition on the encrypted data and returns the result to the user. The user's computer then decrypts the result and combines it with the precomputed carry to produce the final sum. The operations and data flow are depicted in Figure \ref{fig:chen_arc}, which illustrates a secure multi-party computation protocol using the Chen cryptosystem and quantum computing.

\begin{figure}[ht]
	\centering
	\includegraphics[width=0.5\textwidth]{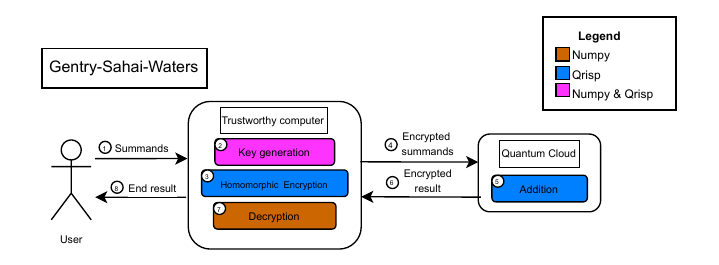}
	\caption{Example architecture of quantum cloud usage for homomorphic addition using the GSW algorithm.} 
	\label{fig:gsw_arc}
\end{figure}

\paragraph{Gentry-Sahai-Waters.} Using the Gentry-Sahai-Waters cryptosystem, the user begins by defining the summands on a trusted computer, which then calculates the keys $\hat{s}$ and $pk$, and the matrices $R$ and $G$ as outlined. The summands are encrypted, and the encrypted data is sent to the quantum cloud. The cloud performs homomorphic addition on the encrypted summands and returns the result to the user's computer. Finally, the computer decrypts the result, revealing the final outcome of the addition operation. The process is illustrated in Figure  \ref{fig:gsw_arc}, in which we can observe the flows for a secure multi-party computation protocol using Gentry-Sahai-Waters cryptosystem and quantum computing.

\begin{figure}[ht]
	\centering
	\includegraphics[width=0.5\textwidth]{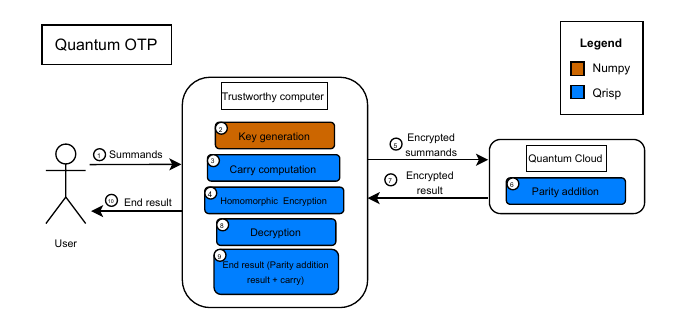}
	\caption{Example architecture of quantum cloud usage for homomorphic addition using the Quantum One Time Pad.} 
  \label{fig:qotp_arc}
\end{figure}

\paragraph{QOTP} Using QOTP, as depicted in Figure \ref{fig:qotp_arc}, the user defines the summands, and the trusted computer generates random keys (a, b, c, d) and computes the carry. The summands are then encrypted using these keys and sent to the quantum cloud. The cloud performs parity addition with the CNOT operator and returns the encrypted result. The computer decrypts this result and combines it with the precomputed carry to produce the final sum.

\section{Implementation}
\label{ch:implementation}
This section discusses the practical aspects of integrating Qrisp with various HE libraries and the challenges encountered. We detail the implementation of the cryptosystems in Python and the setup of the trustworthy computer and quantum cloud using a Flask-based application in a Docker container.

\subsection{Challenges}
Our initial approach combined Qrisp with existing HE libraries, but this revealed compatibility issues between Qrisp's data types and those supported by common HE libraries.

We explored Pyfhel \cite{ibarrondo_pyfhel_2021}, which encodes messages into plaintext objects before encryption. However, Pyfhel only accepts Numpy primitives, not QuantumVariables like QuantumFloats, which are
at the core of Eclipse Qrisp. Additionally, Pyfhel's ciphertexts (PyCtxt) cannot be converted into QuantumVariables, preventing their use in Qrisp.

Similarly, the Concrete library \cite{chillotti_concrete_2020}, an open-source FHE library, tracks and compiles computation graphs for homomorphic evaluation. However, defining functions involving QuantumFloats results in errors, as Concrete only processes certain Python and Numpy functions. Attempts to integrate QuantumVariables led to branching errors, as Qrisp functions are unsupported by Concrete.

These data type incompatibilities were consistent across other HE libraries. Consequently, we implemented the cryptosystems in Python and integrated them directly with Qrisp.

\subsection{Chen}
The key generation process starts by generating Hamming code matrices with \texttt{hamming\_code\_gen}. The generator matrix g is initialized and populated with columns from an identity matrix. A copy of g is made and its transpose is stored in r. For the remaining $N - n$ positions in g, the function randomly assigns $N - n$ columns of the $n \times n$ identity matrix, then swaps the ones and the zeroes. The function also creates a scrambler matrix s, which is a random invertible $n \times n$ matrix with elements zero or one. Additionally, a permutation matrix p is initialized as an identity matrix of size $N \times N$ which is randomly permuted along its columns.

These matrices are then encoded into \texttt{QuantumModulus} arrays to be able to do modular arithmetic later. In \texttt{keygen}, the encryption matrix \texttt{psi} is generated by calculating $\psi = SGP$ and decryption keys \texttt{r}, \texttt{s\_inv} and \texttt{p\_inv} are created.

\begin{minted}{python}
def hamming_code_gen(N, k):
    ...
    return s,g,p,r

def keygen(n):
    k = math.floor(math.log2(n))+1
    N = n + k
    s_prime, g_prime, p_prime, 
		  r_prime = hamming_code_gen(N,k)
    ...
    psi = s@g@p
    return psi, r, s_inv, p_inv
\end{minted}

For the encryption, summands are divided into four-bit segments and encrypted using the encrypt function, which multiplies binary numbers by the encryption matrix \texttt{psi}. This process ensures efficiency by using Hamming(7,4) code, balancing security and computational performance.

\begin{minted}{python}
def encrypt(arr_bin_a, arr_bin_b, psi):
    encrypted = []
    for i in  list(range(0, 
		len(arr_bin_a)-4+1, 4)):
        ...
        cx1 = q_bin_x1@psi
        cx2 = q_bin_x2@psi
        ...
        encrypted.append(cres)   
    return encrypted
\end{minted}

The encrypted summands are sent to the cloud for addition via a \texttt{POST} request. The cloud performs the addition and returns the encrypted sum. The client program reads the response and encodes it into a \texttt{QuantumModulus} array.

The \texttt{decrypt} function uses the matrices \texttt{r}, \texttt{s\_inv} and \texttt{p\_inv} to decrypt the encrypted quantum arrays. The function restores bit significance and combines the decrypted segments to reconstruct the final result using bitwise XOR operations.

\begin{minted}{python}
def decrypt(encrypted, p_inv, r, s_inv):
    result = 0
    i = 0
    for enc in reversed(encrypted):
        dec = enc@p_inv@r@s_inv
        ...
        i += 1
    return result
\end{minted}

\subsection{Gentry-Sahai-Waters}
The \texttt{keygen} function generates a Sophie Germain prime \texttt{q} of bit length \texttt{k}, a user-specified parameter. Sophie Germain primes are valued for their cryptographic robustness. The bit length of \texttt{q} is denoted as $l$, and the dimensions $n$ and $m$ are set to \texttt{k} and $n \times l$, respectively. A secret vector \texttt{s} is randomly initialized with values from $0$ to $q-1$, and the secret key \texttt{t} is \texttt{s} appended with 1, encoded into a quantum array \texttt{t\_q}. A random matrix \texttt{A} of size $(n-1) \times m$ is generated, followed by the formation of matrix \texttt{B} by stacking $-\texttt{A}$ and $\texttt{s} \times \texttt{A}$, then encoded into \texttt{B\_q}.

\begin{minted}{python}
def keygen(k):
    q = generateSophieGermainPrime(k)
    ...
    return k, q, t_q, A, B_q
\end{minted}

The \texttt{encrypt} function encrypts messages using a random binary matrix \texttt{R} and a gadget matrix \texttt{G}, constructed from powers of 2 along the diagonal. The ciphertext is produced by multiplying the public key matrix \texttt{B} with \texttt{R}, and adding the product of the message and \texttt{G}. The encrypted summands are sent to the cloud for addition via a \texttt{POST} request. The cloud performs the addition and returns the encrypted sum. 

\begin{minted}{python}
def encrypt(keys, message):
    R = np.random.randint(2, 
    size=(keys.m, keys.m), dtype=np.int64)
		
    g = 2**np.arange(keys.l)
		
    G = block_diag(*[g for null 
    in range(keys.n)])
    
    return semi_classic_matmul
    (keys.PK, R) + message*G
\end{minted}

The \texttt{decrypt} function recovers the addition result by decoding the encrypted message using the secret key \texttt{t\_q} and the gadget matrix \texttt{G}. The decrypted message is determined by minimizing the distance between the decryption result and the matrix \texttt{sg}.

\begin{minted}{python}
def decrypt(keys, response):
    cres = read_array(response)
    msg = np.dot(sk, cres) % keys.q
    ...
    for mu in mus:
        ...
        if dist < best_dist:
            best_num = mu
            best_dist = dist
    return best_num
\end{minted}

\subsection{QOTP}

The \texttt{encrypt} function applies Pauli-$\mathbf{X}$ and -$\mathbf{Z}$ operations to qubits based on randomly generated integers \texttt{a}, \texttt{b}, \texttt{c}, and \texttt{d}. The encryption keys \texttt{a\_list}, \texttt{b\_list}, \texttt{c\_list}, and \texttt{d\_list} are saved for decryption.

\begin{minted}{python}
def encrypt(j, k):
    ...
    for q1,q2 in zip(j.reg, k.reg):
        a,b,c,d = np.random.randint(2, size = 4)
        ...
        if(a == 1):
            x(q1)
        if(b == 1):
            z(q1)
        if(c == 1):
            x(q2)
        if(d == 1):
            z(q2)
    return a_list, c_list, d_list
\end{minted}

The \texttt{decrypt} function reverses the Pauli-$\mathbf{X}$ and $\mathbf{Z}$ operations applied during encryption using \texttt{a\_list}, \texttt{c\_list}, and \texttt{d\_list}, and combines the result with the previously computed carry.

\begin{minted}{python}
def decrypt(carry, response, a_list, 
c_list, d_list):

    res = read_result(response)
    for i in range(len(res.reg)):
        if a_list[i] != c_list[i]:
            x(res.reg[i])
        if d_list[i] == 1:
            z(res.reg[i])

    return carry + res
\end{minted}

The \texttt{he\_add} function computes the carry using a multi-controlled $\mathbf{X}$ gate, encrypts the summands with \texttt{encrypt}, and sends the data to a cloud service for addition. The cloud returns the encrypted result, which is decrypted and combined with the carry to produce the final sum.

\begin{minted}{python}
def he_add(num1, num2):
    ...
    bit_carry(qnum1, qnum2, carry)
    a_list, c_list, d_list = encrypt(qnum1, qnum2)
    ...
    return decrypt(carry, response, 
		a_list, c_list, d_list)
\end{minted}

\begin{figure}[]
	\centering
	\begin{subfigure}[b]{0.495\textwidth}
		\centering
		\includegraphics[width=0.95\textwidth]{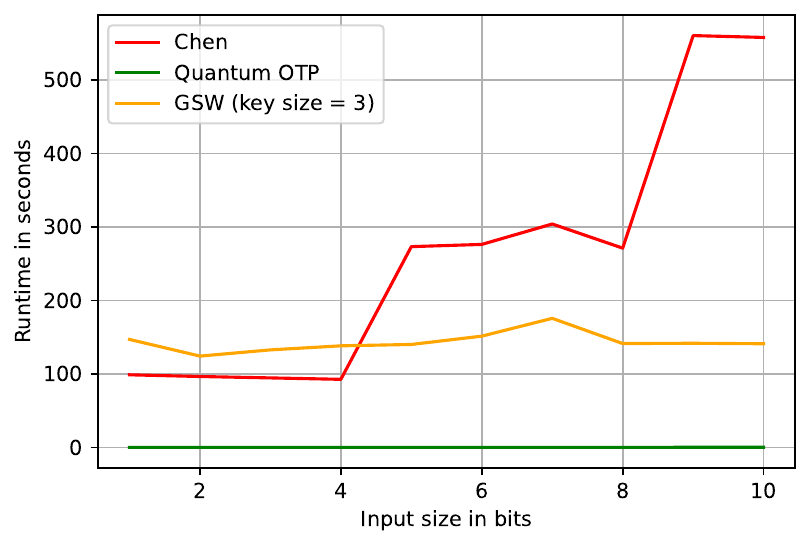}
		\caption{Algorithm Runtimes vs. Input Size}\label{fig:comparison:input}
	\end{subfigure} 
	\hfill
	\begin{subfigure}[b]{0.495\textwidth}
		\centering
		\includegraphics[width=0.98\textwidth]{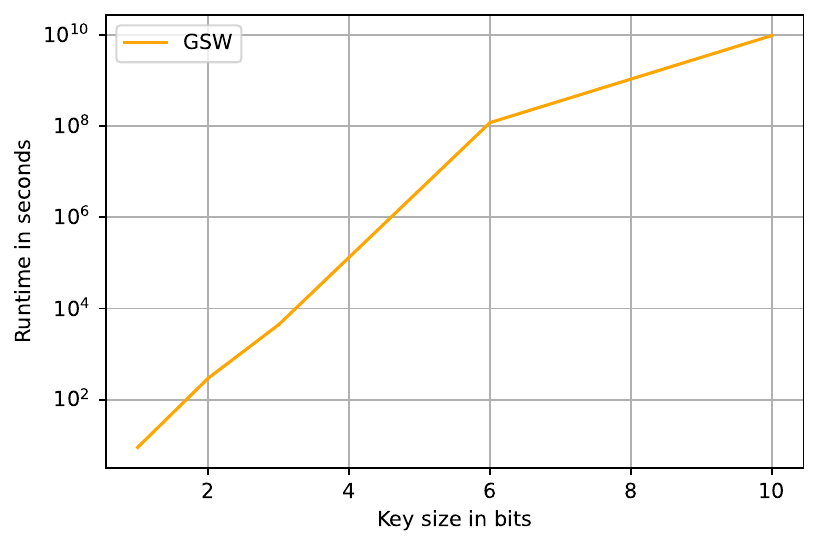}
		\caption{Algorithm Runtimes vs. Key Size (Log Scale)}\label{fig:comparison:key}
	\end{subfigure}
   \caption{Algorithm Runtimes Comparison} 
   \label{fig:comparison}
\end{figure}

\section{Evaluation}
\label{ch:evaluation}
In this section, we evaluate the runtime trends and computational characteristics of each encryption scheme. Our analysis focuses on runtime and memory consumption across varying input and key sizes, aiming to assess computational efficiency and scalability. The specifications of the setup used for the analysis are shown in Table \ref{tab:specifications}.

\subsection{Experimental Setup}

The client script supports various cryptographic systems implemented for this article, interacting with a Flask-based cloud service to perform addition on encrypted integers. It supports various cryptographic systems implemented in this work, sending encrypted values to the cloud and handling the encrypted results.

The script prompts the user to enter two integers and select a cryptosystem. It then generates encryption keys and calls the \texttt{he\_add} function, which encrypts the values, sends them to the cloud service, and receives the encrypted sum. Finally, the result is decrypted and returned to the user.

The cloud service, implemented in Python using Flask and running in a Docker container, performs the addition on encrypted integers. The service has a single \texttt{/process} endpoint that handles \texttt{POST} requests. It processes data based on its format, performs the addition, and returns the result as a \texttt{JSON} response.

\begin{table*}[ht]
\centering
\caption{Specifications of the experimental setup used for evaluating the different HE schemes.}
\label{tab:specifications}
\begin{tabular}{l|l|l}
\multicolumn{1}{c|}{\textbf{Computer Specification}} & \multicolumn{1}{c|}{\textbf{Virtual Machine}} & \multicolumn{1}{c}{\textbf{Software Versions}}  \\ \hline
Processor: Intel(R) Core(TM) i7-1260P 2.10 GHz & JupyterHub version: 4.0.2  & Python 3.11.7 \\
RAM: 32,0 GB & GPUs: 4 x 12GB NVIDIA GPUs & Qrisp 0.4.9 \\
Operating System: Ubuntu 20.04 LTS & RAM: 32,0 GB & Docker 27.0.3     
\end{tabular}
\end{table*}

\subsection{Time Consumption}

Figure \ref{fig:comparison:input} shows that the runtime of all algorithms increases with input size. 
The line graphs compare the runtimes of three algorithms: Chen, Quantum OTP, and GSW (with key size 3).
The QOTP and GSW algorithms maintain relatively Chen's algorithm exhibits periodic increases in runtime, due to its iterative process when summands exceed a certain bit length.

\begin{figure}[]
	\centering
	\begin{subfigure}[b]{0.495\textwidth}
		\centering
		\includegraphics[width=0.95\textwidth]{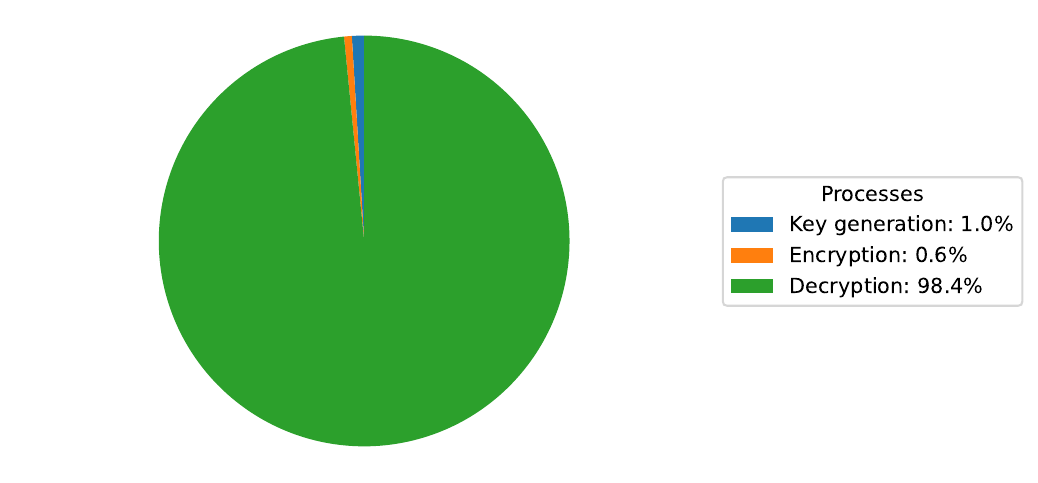}
		\caption{Chen}\label{fig:percentage:Chen}
	\end{subfigure}
	\hfill
	\begin{subfigure}[b]{0.495\textwidth}
		\centering
		\includegraphics[width=0.95\textwidth]{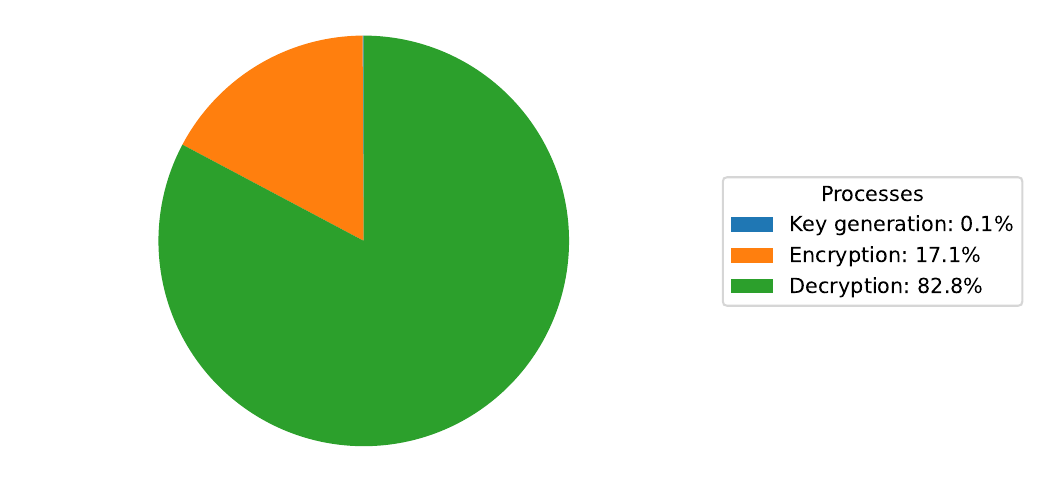}
		\caption{GSW}\label{fig:percentage:gsw}
	\end{subfigure}
	\hfill
	\begin{subfigure}[c]{0.495\textwidth}
		\centering
		\includegraphics[width=0.95\textwidth]{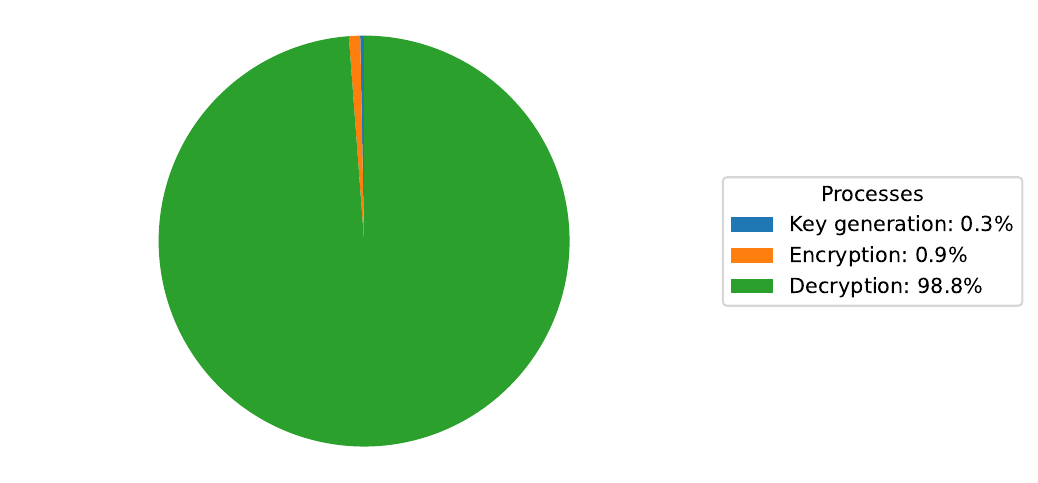}
		\caption{QOTP}\label{fig:percentage:qotp}
	\end{subfigure}
	\caption{Process Contributions to Total Algorithm Runtimes} 
  \label{fig:percentage}
\end{figure}


Figure \ref{fig:percentage:Chen} indicates that decryption in Chen's algorithm is the most time-consuming step. While restructuring to perform decryption only at the end could improve performance, it would disrupt bit significance, leading to incorrect results.

Figures \ref{fig:comparison:input} and \ref{fig:percentage:qotp} show that QOTP is the fastest, with a constant runtime due to direct qubit modification. Its runtime remains constant regardless of key size since it is determined by input size. GSW, however, exhibits exponential growth with key size due to the complexity of its multilinear maps and matrix operations, as shown in Figure \ref{fig:percentage:gsw}. Decryption is the most time-consuming phase for GSW, involving extensive matrix multiplications and distance calculations.

\subsection{Memory Consumption}

\begin{figure}[]
	\centering
	\begin{subfigure}[b]{0.495\textwidth}
		\centering
		\includegraphics[width=0.95\textwidth]{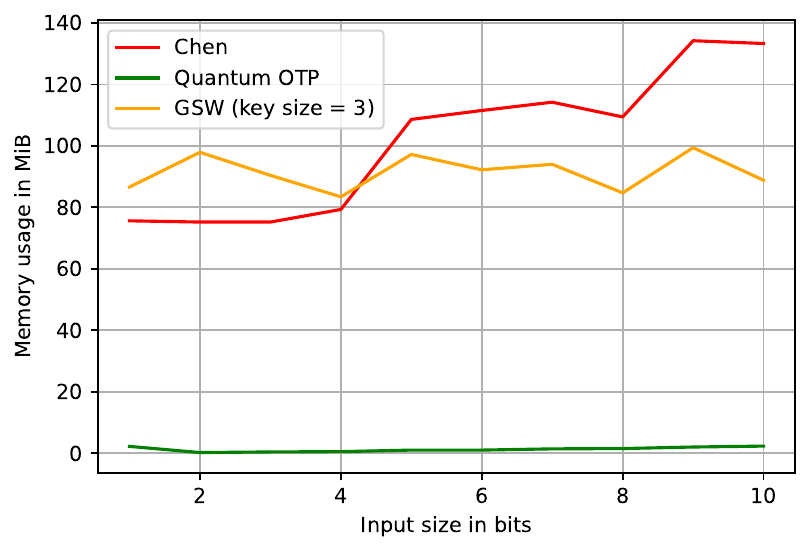}
		\caption{Memory Consumption vs. Input Size}\label{fig:comparison_mem:input}
	\end{subfigure} 
	\hfill
	\begin{subfigure}[b]{0.495\textwidth}
		\centering
		\includegraphics[width=0.98\textwidth]{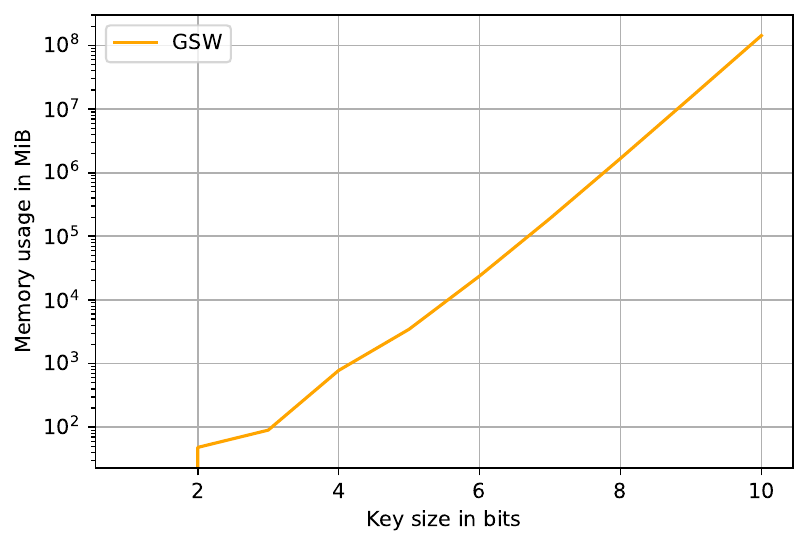}
		\caption{Memory Consumption vs. Key Size (Log Scale)}\label{fig:comparison_mem:key}
	\end{subfigure}
   \caption{Algorithm Memory Consumption Comparison} 
   \label{fig:comparison_mem}
\end{figure}


Figure \ref{fig:comparison_mem:input} shows that memory consumption increases with input size for all algorithms. QOTP and GSW maintain relatively constant memory usage, while Chen’s algorithm shows staggered growth due to binary encodings in multiples of four bits.

Figure \ref{fig:comparison_mem:key} illustrates memory consumption relative to key size. GSW demonstrates a steep increase in memory usage, indicating faster-than-super-linear growth.

\begin{figure}[]
	\centering
	\begin{subfigure}[b]{0.495\textwidth}
		\centering
		\includegraphics[width=0.95\textwidth]{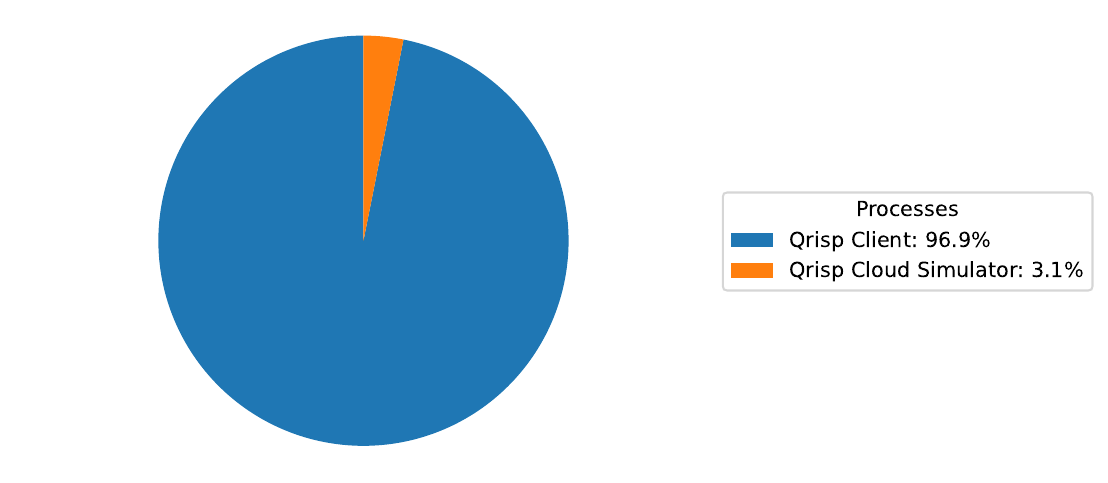}
		\caption{Chen}\label{fig:percentage_mem:Chen}
	\end{subfigure}
	\hfill
	\begin{subfigure}[b]{0.495\textwidth}
		\centering
		\includegraphics[width=0.95\textwidth]{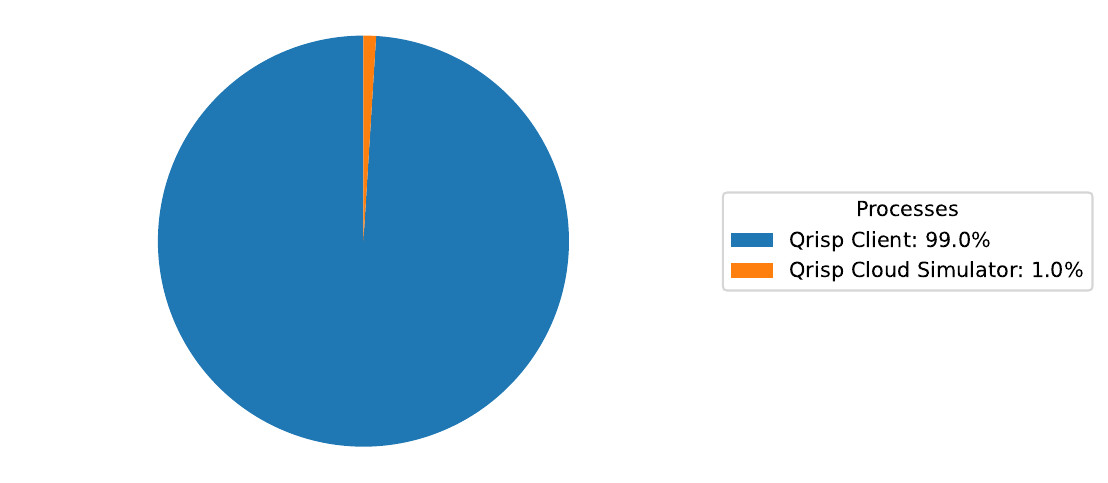}
		\caption{GSW}\label{fig:percentage_mem:gsw}
	\end{subfigure}
	\begin{subfigure}[c]{0.495\textwidth}
		\centering
		\includegraphics[width=0.95\textwidth]{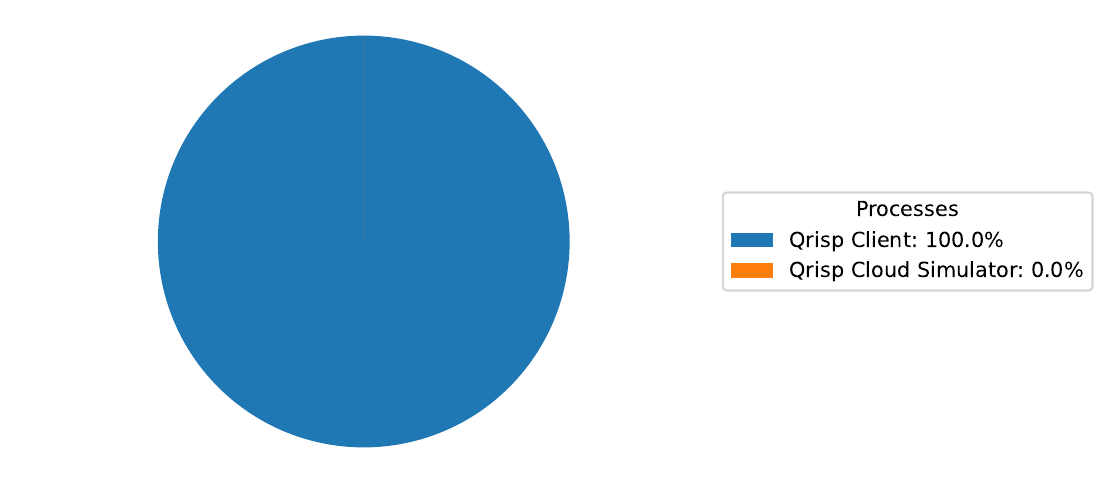}
		\caption{QOTP}\label{fig:percentage_mem:qotp}
	\end{subfigure}
	\caption{Memory Usage Distribution} 
   \label{fig:percentage_mem}
\end{figure}

Figure \ref{fig:percentage_mem} shows that the client program, handling encryption and decryption, consumes the majority of memory across all cryptosystems, with the cloud service handling homomorphic addition consuming less.
\section{Discussion}

The results of our research presented in chapters \ref{ch:implementation} and \ref{ch:evaluation} highlight several key insights to answer our initial research questions.

\subsection{Technical Feasibility}

We implemented  three PQC algorithms based on homomorphic encryption using Qrisp. This demonstrates that quantum-resistant HE is compatible with Qrisp and can be effectively used in larger quantum cloud architectures. Among the PQC algorithms, QOTP stands out due to its simplicity and low overhead in both runtime and memory consumption. Its straightforward design suggests that QOTP is not only well-suited for Qrisp but also adaptable to other quantum computing frameworks.

\subsection{Impact on Future Quantum Cloud Platforms}

Integrating HE into the Qrisp framework could significantly enhance the security and operational capabilities of quantum cloud platforms. However, the performance trade-offs associated with HE must be managed by selecting resource-efficient algorithms. While three of the HE algorithms are resource-intensive, more efficient cryptosystems like QOTP are crucial. Advances in quantum hardware may mitigate the performance impacts of HE, making it more practical for quantum cloud environments. Additionally, implementing HE can help quantum cloud platforms comply with stringent data protection regulations, ensuring data security throughout its lifecycle.

\subsection{Recommendations for Quantum Cloud Security}

Based on current developments in quantum cloud security requirements and the evaluation of quantum-resistant HE implementations, we propose the following recommendations:

\subsubsection{HE at Data Storage and Processing Levels}

Quantum cloud providers should implement HE to secure data at rest, reducing the risk for data breaches. Data should be encrypted before uploading it to the cloud and encryption should be maintained during storage. 
Furthermore, quantum cloud providers should select HE schemes that balance security and performance, such as QOTP, which has low runtime and memory overhead. In addition, quantum cloud providers should implement redundancy and secure backup strategies, ensuring that encrypted backups are stored in multiple locations to prevent data loss from hardware failures or cyber-attacks. HE algorithms should be optimized for the specific quantum cloud environment using hardware acceleration and software optimizations.

\subsubsection{Quantum-Specific Security Measures for Communication and Computation}

Quantum cloud providers should enhance communication security within quantum cloud infrastructures using QKD, which securely exchanges encryption keys based on quantum mechanics principles. In addition, QKD should be combined with quantum-resistant HE at endpoints to secure the entire data pipeline. The trustworthiness of quantum hardware should be enhanced using techniques like QuPUFs to eliminate potential vulnerabilities and increase the level of confidentiality and integrity.

\subsubsection{Access Control and Authentication Mechanisms}

Quantum cloud providers should implement stringent authorization measures to protect data and resources. They should define roles that reflect responsibilities within the quantum cloud infrastructure and regulate user access based on the principle of least privilege. Permissions should be regularly reviewed to ensure they align with users' evolving roles and unnecessary permissions should be revoked. Multi-factor-authentication should be put in place to reduce the risk of unauthorized access.

\subsubsection{Participation in Standardization Efforts}

Participate in standardization efforts to stay ahead of emerging quantum computing threats. Standardization ensures compatibility of quantum-safe cryptographic algorithms across systems and platforms, facilitates seamless integration, and accelerates the adoption of quantum-resistant solutions. Standardized algorithms undergo rigorous evaluation by the global cryptographic community, providing assurance against quantum attacks.

\subsubsection{Robust Incident Response Plans}

Quantum cloud providers should develop dedicated Incident Response Teams (IRT) with experts in quantum computing, cybersecurity, and IT operations. They should outline incident response policies that align with the overall security strategy and regulatory requirements. Furthermore, real-time monitoring tools should be deployed to detect suspicious activities and potential security incidents promptly. Containment strategies should be implemented to limit the spread and impact of incidents. Quantum cloud providers should conduct thorough forensic analysis and document lessons learned to refine response procedures and enhance monitoring capabilities. Regular drills and continuous training should ensure team readiness and adaptability.

\begin{figure*}
	\centering
	\includegraphics[width=\textwidth]{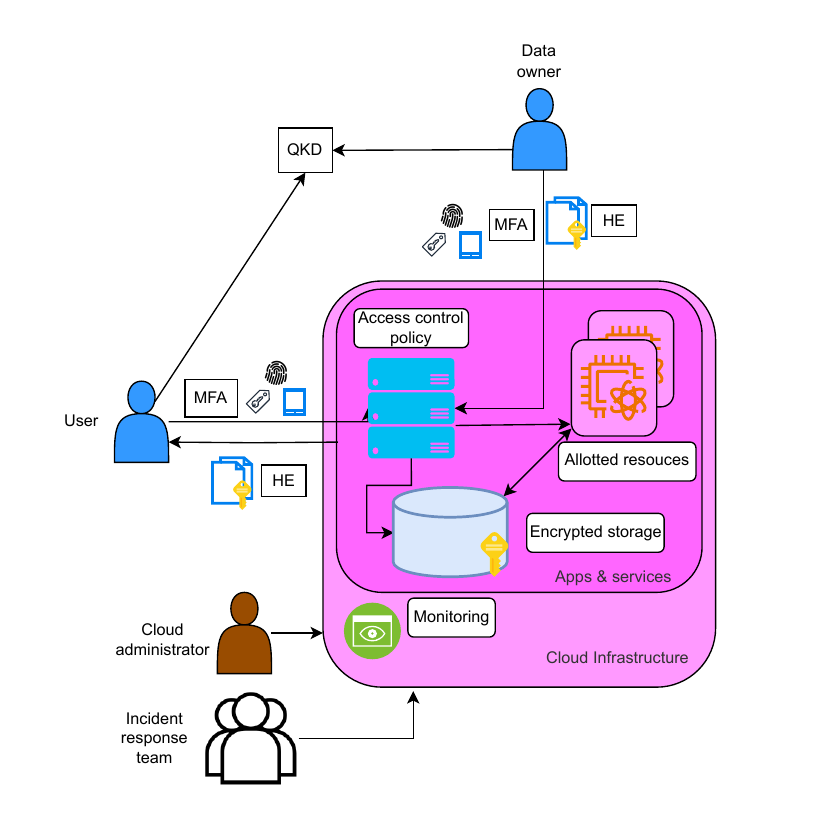}
	\caption{Implementation of our recommendations for quantum cloud security.}
    \label{fig:recs}
\end{figure*}

Figure \ref{fig:recs} illustrates our recommendations. This results in a quantum cloud security architecture incorporating various security measures. Key elements include quantum key distribution, multi-factor authentication, homomorphic encryption, access control policies, encrypted storage, monitoring, and an incident response team. These components work together to protect user data and ensure the security of cloud applications and services.

Users exchange encryption keys using QKD, and data is encrypted with homomorphic encryption both in storage and transit. Users authenticate with MFA and access resources based on stringent access control policies. The system is continuously monitored for potential threats, with a dedicated incident response team addressing any identified issues and implementing containment strategies.

\section{Conclusions and Future Work}
\label{ch:conclusion}

Cloud computing offers on-demand access to shared computing resources, but quantum computing introduces new security challenges. Quantum computers can break traditional cryptographic defenses, necessitating quantum-resistant encryption strategies like homomorphic encryption (HE) to secure data.

This research explored the integration of HE into the Eclipse Qrisp framework, aiming to secure quantum cloud platforms. We addressed challenges in merging HE with Qrisp, assessed technical feasibility, and evaluated the potential impact on future quantum cloud environments.

Our findings indicate that while algorithms like QOTP offer low overhead and simplicity, others like GSW and Chen present performance trade-offs in terms of runtime and memory consumption. The successful implementation of three post-quantum cryptographic (PQC) algorithms within Qrisp—Chen, GSW, and QOTP—demonstrates the technical feasibility of integrating HE with quantum computing frameworks. This integration is crucial for developing secure quantum cloud architectures capable of withstanding quantum threats.

To enhance quantum cloud security, we propose several recommendations: 
\begin{enumerate}
    \item Implementing HE at data storage and processing levels to keep data encrypted throughout its lifecycle
    \item Developing quantum key distribution (QKD) for secure key exchanges resistant to quantum attacks
    \item Enforcing stringent access control and authentication mechanisms to prevent unauthorized access
    \item Participating in standardization efforts to ensure compatibility and robustness against quantum attacks
    \item Maintaining robust incident response plans to promptly address security incidents and improve overall security.
\end{enumerate}

However, the study also highlights limitations which show necessities and possibilities for future work:
\begin{itemize}
    \item \textbf{Focus on Homomorphic Encryption}: While HE provides robust data security, it is only one of many cryptographic techniques. Future research should explore alternative cryptographic methods, such as lattice-based cryptography and hash-based signatures, to evaluate their applicability and effectiveness in quantum cloud settings. Investigating these alternatives could provide a more comprehensive understanding of the various approaches to enhancing quantum cloud security.
    \item \textbf{Quantum Cloud Use Cases}: This work focuses on specific use cases within the quantum cloud, primarily HE addition. However, quantum clouds can support a wide range of applications, including machine learning, optimization, and complex simulations. Future research should examine the security requirements and performance implications of HE in these diverse applications to provide a broader understanding of its effectiveness across different quantum cloud use cases.
    \item \textbf{Performance Metrics}: The performance metrics used in this evaluation (key generation time, encryption and decryption time, and memory consumption) may not fully capture the real-world complexities of quantum cloud environments. Future work should include stress testing under various operational conditions, such as network latency and multi-tenant environments, to gain a more accurate understanding of the practical implications and scalability of the proposed security solutions. Long-term performance monitoring is also essential to assess the stability and efficiency of HE schemes over time.
    \item \textbf{Evolving Threat Landscape}: The threat landscape for quantum computing is rapidly evolving, with new vulnerabilities and attack vectors emerging. The security measures evaluated in this work are based on current threats, but as quantum technology advances, so will the sophistication of potential attacks. Continuous research and adaptive security strategies are essential to stay ahead of these emerging threats. This includes developing quantum-resistant cryptographic algorithms, implementing robust security protocols, and establishing a dynamic security framework that can quickly adapt to new threats. Regular updates to cryptographic standards and ongoing threat assessments will be crucial.
\end{itemize}



In conclusion, while integrating HE into Eclipse Qrisp shows potential, careful consideration of performance trade-offs and ongoing research into evolving threats are essential for the future development of secure quantum cloud platforms.

\section*{Acknowledgements}
This work was funded by the Federal Ministry for Economic Affairs and Climate Action (German: Bundesministerium für Wirtschaft und Klimaschutz) under the project SeQuenC (01MQ22009C). The authors are responsible for the content of this publication.

\bibliographystyle{unsrt}
\bibliography{bibliography}

\end{document}